\documentclass[prl, reprint, twocolumn, aps, superscriptaddress]{revtex4-1}
\usepackage{graphicx}
\usepackage{dcolumn}
\usepackage{bm}
\usepackage{dcolumn}
\usepackage{color,soul}
\definecolor{purple}{RGB}{127,0, 255}

\usepackage{ragged2e}
\usepackage{amssymb}
\usepackage{amsmath}
\begin{document}
\title{Enhanced Superconductivity in Monolayer $T_d$-MoTe$_2$ with Tilted Ising Spin Texture}
\author{Daniel A. Rhodes}
\email{darhodes@wisc.edu}
\affiliation{Department of Mechanical Engineering, Columbia University, New York, NY 10027}
\author{Apoorv Jindal}
\affiliation{Department of Physics, Columbia University, New York, NY, 10027}
\author{Noah F. Yuan}
\affiliation{Department of Physics, Massachusetts Institute of Technology, Cambridge, MA 02142}
\author{Younghun Jung}
\affiliation{Department of Mechanical Engineering, Columbia University, New York, NY 10027}
\author{Abhinandan Antony}
\affiliation{Department of Mechanical Engineering, Columbia University, New York, NY 10027}
\author{Hua Wang}
\affiliation{Department of Materials Science and Engineering, Texas A\&M University, College Station, Texas 77840}
\author{Bumho Kim}
\affiliation{Department of Mechanical Engineering, Columbia University, New York, NY 10027}
\author{Yu-che Chiu}
\affiliation{Department of Physics and National High Magnetic Field Laboratory, Florida State University, Tallahassee, FL 32306}
\author{Takashi Taniguchi}
\affiliation{National Institute for Materials Science, 1-1 Namiki, Tsukuba 305-0044, Japan}
\author{Kenji Watanabe}
\affiliation{National Institute for Materials Science, 1-1 Namiki, Tsukuba 305-0044, Japan}
\author{Katayun Barmak}
\affiliation{Department of Applied Physics and Applied Mathematics, Columbia University, New York, NY 10027}
\author{Luis Balicas}
\affiliation{Department of Physics and National High Magnetic Field Laboratory, Florida State University, Tallahassee, FL 32306}
\author{Cory R. Dean}
\affiliation{Department of Physics, Columbia University, New York, NY, 10027}
\author{Xiaofeng Qian}
\affiliation{Department of Materials Science and Engineering, Texas A\&M University, College Station, Texas 77840}
\author{Liang Fu}
\affiliation{Department of Physics, Massachusetts Institute of Technology, Cambridge, MA 02142}
\author{Abhay N. Pasupathy}
\email{apn2108@columbia.edu}
\affiliation{Department of Physics, Columbia University, New York, NY, 10027}
\author{James Hone}
\email{jh228@columbia.edu}
\affiliation{Department of Mechanical Engineering, Columbia University, New York, NY 10027}

\begin{abstract}
Crystalline two-dimensional (2D) superconductors (SCs) with low carrier density are an exciting new class of materials in which electrostatic gating can tune superconductivity, carrier-carrier interactions play a strong role, and bulk properties directly reflect the topology of the Fermi surface. Here we report the dramatic enhancement of superconductivity with decreasing thickness in semimetallic $T_d$-MoTe$_2$, with critical temperature reaching up to 7.6 K for monolayers, a sixty-fold enhancement as compared to the bulk. We show that monolayers possess a similar electronic structure and density of states (DOS) as the bulk, indicating that electronic interactions play a strong role in the enhanced superconductivity. Reflecting the low carrier density, the critical temperature, magnetic field, and current density are all tunable by an applied gate voltage. The response to high in-plane magnetic fields is distinct from that of 2D SCs such as $2H$-NbSe$_2$ and reflects the canted spin texture of the electron pockets.
\end{abstract}
\maketitle

\indent In the 2D limit, superconductivity can differ strongly from the bulk due to spatial confinement and increased interactions, and can be tuned electrostatically, opening up new possibilities for device applications\cite{Super2DRev}. The 2D limit of superconductivity has been extensively studied in metal and metal-oxide films, but in these systems disorder plays a strong role and interaction with the substrate or capping layers through strain or charge transfer can strongly alter superconducting behavior\cite{FeSeSiC,FeSe109, Super2DRev}. Recently, exfoliated 2D metallic SCs such as NbSe$_2$\cite{NbSe2Ising} and TaS$_2$\cite{barrera_tuning_2018}, as well as 2D semiconductors (i.e. MoS$_2$) with high induced carrier density through ionic liquid gating\cite{JTYeMoS2}, have provided a new opportunity to study 2D superconductivity in crystalline systems with weak substrate interaction. This has resulted in the observation of many new phenomena, such as enhanced upper critical fields from strong out-of-plane spin-orbit coupling (SOC), enhanced $T_c$ as one approaches the monolayer limit\cite{talantsev_origin_2017}, and even electrostatic control over the dissipation of vortices\cite{benyamini2019fragility,benyamini2019blockade}. Very recently, superconductivity  has been discovered in twisted bilayer graphene ($T_\textrm{c}$ = 2 K) and monolayer $T_d$-WTe$_2$ ($T_\textrm{c}$ = 700 mK), at low carrier densities ($<10^{13}/$cm$^2$ induced by an electrostatic gate). The strong carrier-carrier interactions and facile tuning by conventional gates in this regime has generated intense interest and motivates the search for additional low-density 2D superconductors.\\ 
\indent Here we explore the superconducting properties of $T_d$-MoTe$_2$ in the 2D limit. In the bulk, $T_d$-MoTe$_2$ is a type-II Weyl semimetal\cite{TypeIIWSM} with a carrier density of $6\times10^{19}$ cm$^{-3}$ \cite{zhou_hall_2016}, and $T_\textrm{c}$ of 120 mK, which can be enhanced by doping or applied pressure\cite{mandal_enhancement_2018,YQiMoTe2SC}. To date, no work has explored $T_d$-MoTe$_2$ in the intrinsic (low-density) 2D limit: few-layer films grown by chemical vapor deposition (CVD) exhibit superconductivity with $T_\textrm{c}$ up to 3 K\cite{CVDSuper}, but are highly doped, with carrier densities of $\sim 3\times10^{14}/\textrm{cm}^2$, \cite{CVDSuper,CVDMoTe2}.\\
\indent In this study, we utilize high-quality single crystals synthesized by a self-flux technique; recent studies indicate that this high quality is maintained after mechanical exfoliation\cite{edelberg_approaching_2019}. Thin flakes were exfoliated inside an inert-atmosphere glove box and encapsulated between $\sim$30 nm thick hexagonal boron nitride ($h$BN) crystals to reduce environmental disorder and provide protection from degradation in air\cite{DennisMoTe2}. Hermetically sealed electrical contacts were obtained by embedding metal within the $h$BN \cite{ViaEvan} or by encapsulating pre-patterned contacts (see Supplementary for details). As we detail below, this approach maintains intrinsic low carrier density and allows us to access the clean limit, where the normal-state mean free path exceeds the superconducting coherence length.\\ 
%
%
\begin{figure}
	\includegraphics[width=\linewidth]{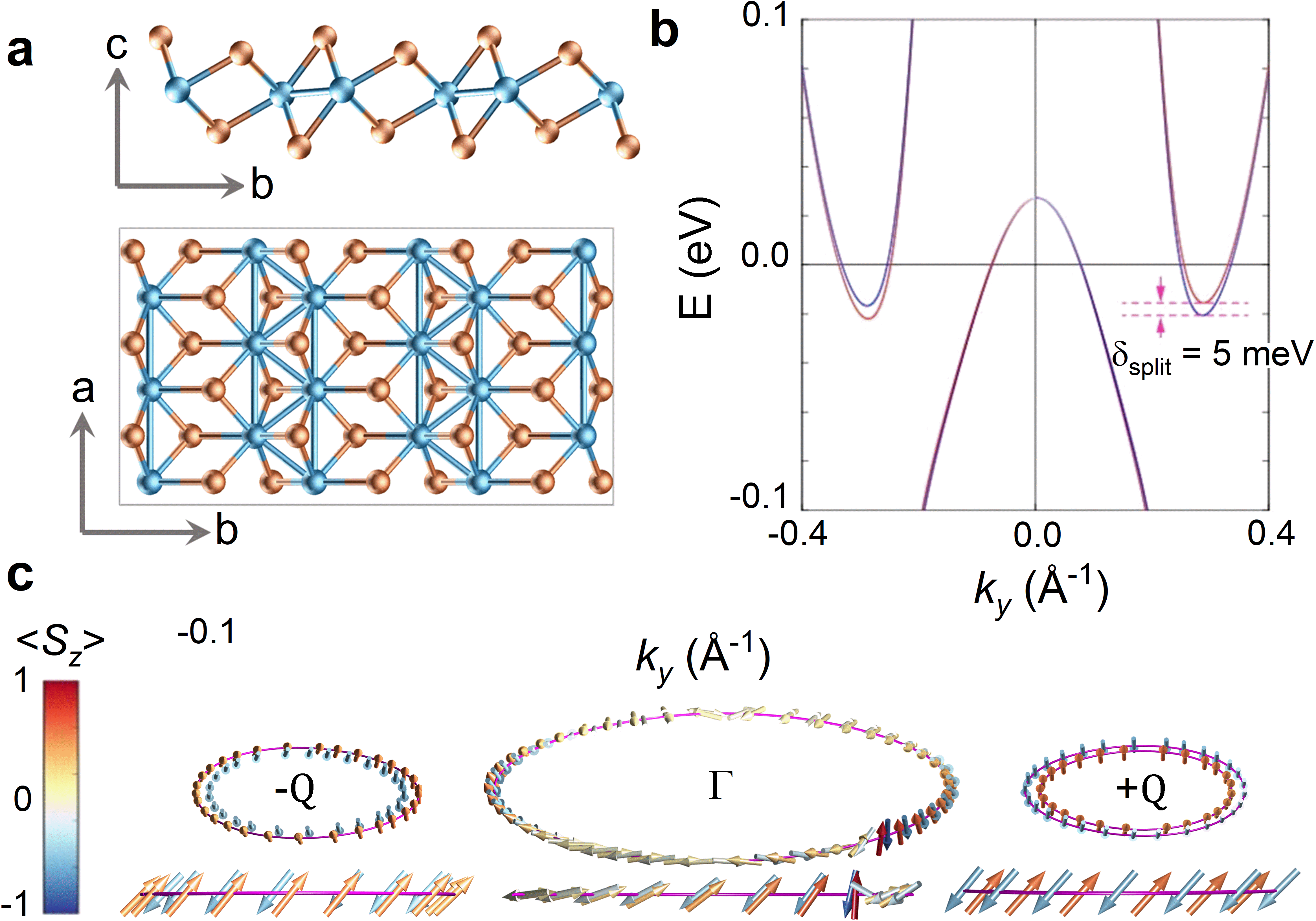}
	\caption{Crystal structure and electronic band structure of $T_d$-MoTe$_2$. (a), Crystal structure of the monolayer.  Blue atoms are Mo, red Te.(b), Calculated electronic band structure.(c), Calculated spin texture, with the spin direction indicated by arrows. The bottom-most cartoon depicts the spin texture along the $z-k_x$ plane.}
\label{Fig. 1}
\end{figure}
\indent  The crystal structure of monolayer $T_d$-MoTe$_2$ (1L-MoTe$_2$) is isostructural to that of WTe$_2$ (Fig. 1a). Previous studies have shown that bulk crystals transition to the $T_d$ phase upon cooling below 240 K,\cite{Bullett} whereas for thin samples the $T_d$ phase is stabilized to above room temperature\cite{TsenMoTe2}. We note that the $T^\prime$ and $T_d$ phases differ primarily in layer stacking, and are identical for monolayers. Figure 1b shows the calculated electronic bandstructure, with one hole pocket at the $\Gamma$ point with a carrier density of $n$ = 1.2$\times10^{13}/$cm$^2$ and two electron pockets ($n$ = 0.6$\times10^{13}/$cm$^2$) at either side of the $\Gamma$ point, denoted as the $\pm Q$ pockets. With a small out-of-plane electric field, inversion symmetry is broken and significant spin-orbit coupling (SOC) develops\cite{shi_symmetry_2019}. For this reason, and in reasonable agreement with our data as discussed below, the bandstructure was calculated under the assumption of an applied, out-of-plane electric field of 0.1 V/nm. With this SOC, the $\Gamma$ pocket is nearly spin degenerate, and the $\pm Q$ pockets exhibit a spin-splitting of $\sim 5$ meV. In both pockets, the spins are tilted rather than being locked entirely out-of-plane like in $2H$-phase materials such as NbSe$_2$. For the $\Gamma$ pocket, the spin orientation depends strongly on the momentum orientation, while for the electron pockets the spin is tilted in the direction of the $b$-axis, independent of momentum (Fig. 1c).  Conclusive experimental verification of this bandstructure is lacking. One report indicates a potential bandgap in few-layered $T_d$-MoTe$_2$\cite{keum_bandgap_2015}, while two recent angle-resolved photoemission spectroscopy (ARPES) studies on 1L-MoTe$_2$ show conflicting results: semimetallic behavior with large band overlap for 1L-MoTe$_2$ grown on graphene\cite{MonoArpes}, and weak overlap with a potential gap opening for 1L-MoTe$_2$ exfoliated on gold\cite{pawlik_thickness_2018}.\\ 
\begin{figure*}[t] 
	\includegraphics[width=.9\linewidth]{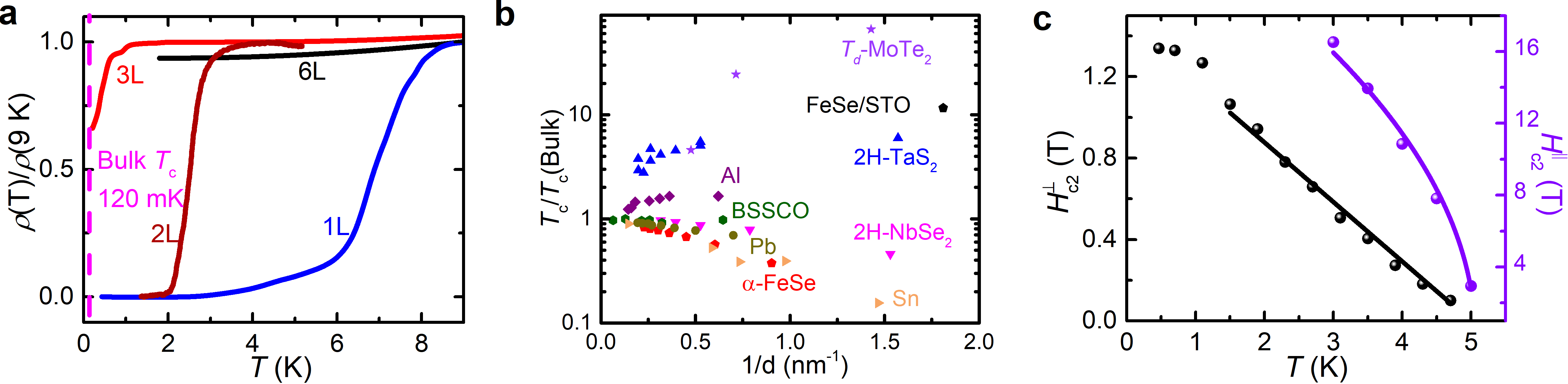}
	\caption{Temperature dependent transport properties.(a), Resistivity as a function of the temperature for 1L, 2L, 3L, and 6L samples. The dashed line indicates $T_\textrm{c}$ for bulk $T_d$-MoTe$_2$.(b) Temperature dependence of $H^\perp_\textrm{c2}$ and $H^\parallel_\textrm{c2}$ to the $ab$ plane for S2. Solid lines are fits to GL theory.(c) $T_\textrm{c}$ as a function of the inverse sample thickness, 1/d, in $T_d$-MoTe$_2$, BSCCO\cite{bsccoyuanbo, bsccophilip}, $\alpha$-FeSe\cite{FeSeSiC}, FeSe on STO\cite{FeSe109}, Sn\cite{liao2018superconductivity}, Al\cite{adams2017zeeman}, Pb\cite{ozer2006growth}, $2H$-TaS$_2$\cite{barrera_tuning_2018}, and $2H$-NbSe$_2$\cite{NbSe2Ising}.} 
\label{Fig. 2}
\end{figure*}
\indent Supplementary figure 2 shows the temperature-dependent resistivity of bulk crystals and $h$BN-encapsulated samples with 1, 2, 3, and 6 layers. The bulk crystals show metallic behavior with residual resistance ratios of up to 2000, an almost two orders of magnitude improvement over recent reports\cite{YQiMoTe2SC}, attesting to the high sample quality. The crystals show the expected $T'$-$T_d$ phase transition at 240 K, whereas the monolayer (1L), 2L and 6L samples do not, consistent with previous studies which indicate that the $T_d$ phase persists above room temperature\cite{TsenMoTe2}. Consistent with the calculated bandstructure, these samples remain metallic down to low temperature, ruling out the presence of a bandgap. Super-linear normal-state magnetoresistance and low normal-state Hall coefficient (Supplementary Fig. 3) provide further evidence that 1L-MoTe$_2$, like the bulk\cite{zhou_hall_2016}, is a nearly charge compensated semimetal. Shubnikov-de Haas (SdH) oscillations in 2L and 3L samples also indicate a carrier density of $\sim 1-2\times10^{13}/\textrm{cm}^2$ (Supplementary Fig. 4). Therefore, we conclude that the electronic structure, carrier density, and DOS do not change substantially from the bulk to the monolayer.\\ 
\indent The central result of this paper is shown in Fig. 2a. Upon cooling, 1L, 2L, and 3L $T_d$-MoTe$_2$ show a strong enhancement of superconductivity compared to the bulk. The monolayer sample shown (S1) has a $T_\textrm{c}$ of $7.6$ K, almost two orders of magnitude larger than the bulk $T_\textrm{c}$. A second sample (S2) studied below shows $T_\textrm{c} = 5$ K. The 2L and 3L samples show $T_\textrm{c}$ of 2.5 K and $\sim$0.5 K, respectively. In contrast, the 6L sample shows no superconducting behavior down to 20 mK (see Supplementary). As discussed above, CVD-grown samples show very different thickness-dependence and high doping suggesting a different origin of enhanced superconductivity.\\
\indent In conventional SCs many factors tend to suppress superconductivity in the 2D limit. For instance, Ginzburg-Landau (GL) theory predicts suppression of superconductivity when film thickness is below the bulk coherence length\cite{simonin_surface_1986}, as seen in other crystalline 2D SCs (e.g. $2H$-NbSe$_2$\cite{NbSe2Ising} and 1UC FeSe on bilayer graphene\cite{FeSeSiC}). In addition, repulsive electron-electron interactions, represented by the screened Coulomb pseudopotential $\mu^*$ in Eliashberg theory, increase as materials approach the 2D limit. Increasing $\mu^*$ suppresses $T_\textrm{c}$ by decreasing the effective pairing interaction, given approximately as $N(0)V=\frac{\lambda-\mu^*}{1+\lambda}$,\cite{carbotte_properties_1990} where $\lambda$ represents the retarded pairing interaction. Thus, the absence of superconductivity in 6L-MoTe$_2$ represents the expected result, whereas its enhancement in 3L, 2L, and 1L samples is surprising.\\ 
\indent Enhancement of $T_\textrm{c}$ in the 2D limit (without induced doping or other extrinsic factors) is extremely rare, with a few prominent examples shown in Fig. 2b. $T_\textrm{c}$ notably increases by an order of magnitude\cite{FeSe109} for monolayer FeSe grown on strontium titanate (STO). While the mechanism not fully understood\cite{huang_electronic_2017}, the enhancement is clearly a result of substrate interaction, and the opposite trend is seen for FeSe on weakly interacting substrates\cite{FeSeSiC}. $2H$-TaS$_2$ shows increasing $T_\textrm{c}$ for thicknesses below the coherence length\cite{barrera_tuning_2018}, with a four-fold enhancement in the monolayer\cite{talantsev_origin_2017}. Potential mechanisms include an increase in the DOS \textit{via} suppression of a competing charge-density wave\cite{yang_enhanced_2018}, decreased interlayer coupling\cite{barrera_tuning_2018}, and the formation of a second superconducting band \cite{talantsev_origin_2017}. As noted above, twisted bilayer graphene and WTe$_2$ are superconducting upon injection of charge from an external gate.\\
\indent 1L-MoTe$_2$ is, therefore, the only known example of a 2D SC which shows large enhancement of $T_\textrm{c}$ while maintaining a similar carrier density and DOS to that of the bulk. The increase in $T_\textrm{c}$ must therefore come from an increase in the pairing interaction. Such an increase does not seem to be consistent with phonon-mediated superconductivity: in van der Waals materials, in-plane phonons do not significantly change for monolayers, and as discussed above $\mu^*$ is expected to be largest for monolayers. In fact, there is already significant evidence that superconductivity in bulk $T_d$-MoTe$_2$ is electronically mediated\cite{MyMoTe2,Zurab,1TMoTe2SC}. We postulate that the change in the Coulomb interaction with dimensionality may account for the enhanced $T_\textrm{c}$ via spin fluctuations, although more work is necessary to conclusively establish the mechanism of superconductivity in this material.\\
\begin{figure*}[t] 
    \centering
	\includegraphics[width= .85\linewidth]{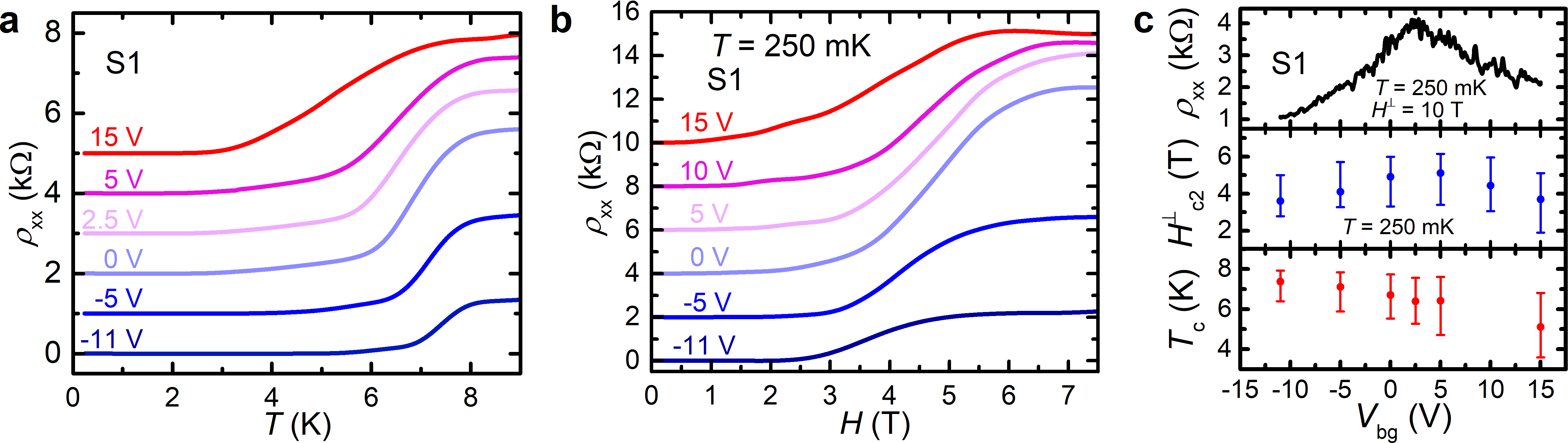}
	\caption{Gate tunability of the superconducting transition. (a), Temperature dependence of $\rho_\textrm{xx}$ from $V_\textrm{bg}$ = -11 to 15 V for sample S1.(b) Magnetic field dependence of $\rho_\textrm{xx}$ with varying gate voltage. All curves are vertically displaced by 1 k$\Omega$, (a), and 2k$\Omega$, (b), for clarity.(c) (top) Normal-state $\rho_\textrm{xx}$ ($H^\perp$ = 10 T); $H^\perp_\textrm{c2}$ (middle) ; (bottom) $T_\textrm{c}$ as a function of the gate voltage and for $R = 50$\%$R_\textrm{N}$. Error bars represent values of $H^\perp_\textrm{c2}$ and $T_\textrm{c}$ at $R = 90$\%$R_\textrm{N}$ (upper) and $R = 10$\%$R_\textrm{N}$ (lower).}
\label{Fig. 3}
\end{figure*}
%
\indent Next we characterize the superconducting phase diagram of monolayer sample S2 under applied magnetic fields. Figure 2c shows the measured perpendicular ($H^{\perp}_\textrm{c2}$) and parallel ($H_\textrm{c2}^\parallel$) critical fields as a function of temperature, defined at 90$\%$ of the normal-state resistivity. $H^\perp_\textrm{c2}$ reaches 1.5 T at 0 K, and decreases linearly with increasing temperature above 1 K, consistent with the 2D GL equation for fields out of plane: $\mu_0H^\bot_\textrm{c2}=\frac{\Phi_0}{2\pi\xi_0^2}(1-T/T_\textrm{c})$, where $\Phi_0$ is the flux quantum, $\mu_0$ the permeability of free space, and $\xi_0$ the in-plane coherence length\cite{Tinkham}. A linear fit (solid line) yields $\xi_0= 13.9$ nm (16 nm for $\mu_0H^\bot_\textrm{c2}=\frac{\Phi_0}{2\pi\xi_0^2}$ at 350 $mK$), roughly half the bulk coherence length\cite{Zurab}. This value is much smaller than the lower bound for the electronic mean free path, $l = 217$ nm (see Supplementary Table 1). Therefore, whereas recent studies of 1L-WTe$_2$ have been in the dirty limit ($\xi_0 >> l$), this work explores the regime where $\xi_0 < l$ and the effects of spin-orbit scattering (SOS) on the superconductivity can be ignored. At fields up to 16 T, $H_\textrm{c2}^\parallel$ follows a square root temperature dependence, and critical field well beyond the Pauli limit ($H_\text{p} = 1.84 T_\textrm{c}= 9.2$ T), consistent with other 2D SCs with strong spin-orbit coupling such as NbSe$_2$ \cite{NbSe2Ising}.\\ 
%
\indent We now examine the gate-tunability of superconductivity. While both S1 and S2 exhibit changes in the superconducting behavior as a function of electrostatic doping, S2 utilizes the Si/SiO$_2$ substrate as its backgate, which introduces static doping from trapped charges in the oxide. Therefore, Fig. 3 shows data for S1, which utilizes a graphite back-gate that minimizes static doping. From the thickness and dielectric constant of \textit{h}BN, we estimate that the induced carrier density is $0.5 \times 10^{12}/$cm$^2$ per applied volt, such that the net carrier density is modified by $1.3 \times 10^{13}/$cm$^2$ over the gate voltage range shown. The normal-state resistivity (Fig. 3c, top) shows a peak near 3 V, consistent with near-charge compensation, while the tunability away from this peak is consistent with a band overlap of order $10^{12}-10^{13}/$cm$^2$. Figures 3a and 3b show gate tuning of the superconducting transition with temperature (Fig. 3a) and out-of-plane magnetic field (Fig. 3b). From these, we extract $T_\textrm{c}$ and $H^\perp_\textrm{c2}$, and plot these values as a function of $V_\textrm{bg}$ in Fig. 3c. For this sample (S1), $T_\textrm{c}$ increases as more hole carriers are injected: a maximum $T_\textrm{c}$ of $\sim8 K$ is observed in the hole dominated region, and a minimum $T_\textrm{c}$ of $\sim5$ K in the electron dominated region. In S2, we see the reverse behavior (see Supplementary Fig. 5), namely a decrease in $T_\textrm{c}$ as the hole concentration is increased as well as a concurrent increase in the normal-state resistance, suggesting that S2 is overdoped with holes. Comparing the electrical response of these two samples and based on the sheet resistance of S2, we postulate the existence of a superconducting dome which relies on both the hole and electron pockets and likely peaks around $10^{14}$/cm$^2$ hole density.\\ 
\indent Whereas $T_\textrm{c}$ decreases uniformly in S1 with increasing $V_\textrm{bg}$, $H_\textrm{c2}$ peaks concurrently with the normal-state resistivity. The reason for the different gate-dependence of $T_\textrm{c}$ and $H_\textrm{c2}$ is unclear, but may be related to variation in mean free path and/or coherence length near neutrality, such that the system moves toward the dirty limit\cite{Tinkham}. Finally, we note that for both magnetic field and temperature the superconducting transition widens as we dope the system with more electrons for both S1 and S2. Direct probes of the crystal structure or electronic band structure at low temperature, such as Raman or ARPES, with gating may shed more light on the origin of this effect.\\
%
\begin{figure*}[t] 
	\includegraphics[width = \linewidth]{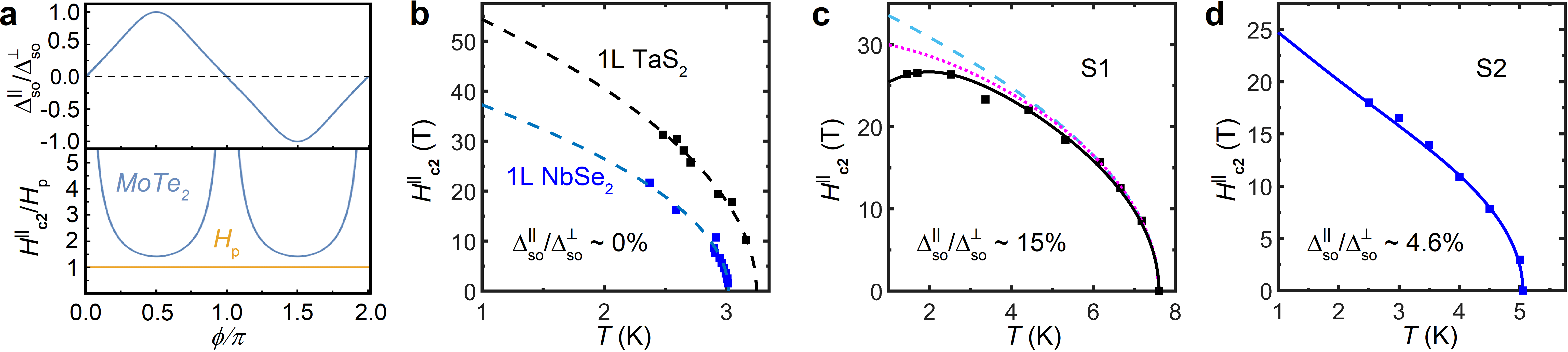}
	\caption{Effects of SOC on superconductivity.(a) (top panel), Calculated ratio between in- and out-of-plane SOC as a function of the in-plane azimuthal angle, $\phi$. (bottom panel), Calculated enhancement of the in-plane upper critical field as a function of $\phi$. (b), Measured $H^\parallel_\textrm{c2}$ for fields parallel to the \emph{ab}-plane in 1L $2H$-TaS$_2$ and $2H$-NbSe$_2$.(c), $H^\parallel_\textrm{c2}$ \textit{vs} $T_\textrm{c}$ for S1. The dotted magenta line is a fit to KLB theory; (d), $H^\parallel_\textrm{c2}$ \textit{vs} $T_\textrm{c}$ for S2. The solid lines are fits to the theory described in the text. Dashed lines are fits to GL theory. (b) adapted from ref.\cite{NbSe2Ising} and ref.\cite{barrera_tuning_2018}.} 
\label{Fig. 4}
\end{figure*}
\indent As discussed above, in these samples the normal-state mean free path exceeds the coherence length, such that SOS plays a minimal role. In this regime, underlying spin texture of the Fermi surface should modify the response to high in-plane magnetic fields. The spin texture is determined by the momentum-dependent SOC vector, $\bm{g_k}$. At the $\Gamma$ point, this quantity must vanish (to leading order in momentum) due to its odd parity under time-reversal symmetry, and higher-order effects lead to a strongly momentum-dependent spin texture for $T_d$-MoTe$_2$ (Fig. 1c). On the other hand, at non-time-reversal-invariant points, the SOC vector, to leading order, becomes a nonzero constant Zeeman field, denoted as $\bm{g}$, which couples to only one component of electron spin and is independent of momentum. To recover overall time-reversal symmetry, at the time-reversal partner $-\bm{k}$, the SOC vector becomes $-\bm{g}$. This is known as Ising SOC\cite{JTYeMoS2, NbSe2Ising, barrera_tuning_2018}, where the overall spin texture resulting from this SOC is determined by the point group symmetry. In $2H$-TMDs, the symmetry of the spin-split hole pockets locks the spins directly in the out-of-plane direction. 1L-MoTe$_2$, on the other hand, has an Ising SOC vector which is \textit{tilted} with respect to the out-of-plane direction (Figure 1a), and has a polar angle of $\sim 45^\circ$ due to the low symmetry at the $Q$ points\cite{shi_symmetry_2019}.\\
\indent The effects of Ising spin texture on $H^\parallel_\textrm{c2}$ for $2H$-TMDs has been described in previous literature by a linearized gap equation which takes into account out-of-plane SOC\cite{NbSe2Ising,barrera_tuning_2018}. This model accounts for both the enhancement of $H^\parallel_\textrm{c2}$ relative to the Pauli limit, and its square-root temperature dependence $H^\parallel_\textrm{c2}(T) = H_0(1-T_\textrm{c}/T_\textrm{c0})^{1/2}$, similar to standard GL theory\cite{NbSe2Ising}. Here we extend the theory to the tilted Ising structure of MoTe$_2$. Given an in-plane magnetic field $\bm{H} = H(\cos\phi, \sin\phi,0)$, we compute $\Delta^\parallel_\textrm{so} = \bm{g}\cdot \bm{\hat{H}}$ and $\Delta^\perp_\textrm{so} = |\bm{g} \times \bm{\hat{H}}|$, which both depend on the in-plane angular direction, $\phi$, of the applied magnetic field, Fig. 4a. With the spin tilted at $45^{\circ}$, the ratio $|\Delta^\parallel_\textrm{so}/\Delta^\perp_\textrm{so}|$ can vary from 0 to 1, depending on the magnetic field direction. For $\phi=0,\pi$ $|\Delta^\parallel_\textrm{so}/\Delta^\perp_\textrm{so}|\sim0$, and the behavior is identical to that of the $2H$-TMDs. In contrast, when $|\Delta^\parallel_\textrm{so}/\Delta^\perp_\textrm{so}|\neq0$, the solution to this linearized gap equation is instead a digamma function (see Supplementary), where the $H^\parallel_\textrm{c2}(T)$ dependence on temperature exhibits a peak as $T$ approaches 0 K. As $|\Delta^\parallel_\textrm{so}/\Delta^\perp_\textrm{so}$ is increased, this peak occurs at lower fields and shifts towards $T_\textrm{c}$. Thus we expect that both the magnitude and temperature dependence of $H^\parallel_\textrm{c2}$ will vary with $\phi$. In samples with unknown orientation such as used in this study, this should lead to variability not seen in the $2H$-TMDs.\\
\indent Figure 4b-d shows the measured $H^\parallel_\textrm{c2}$ for two model $2H$-TMDs, NbSe$_2$ and TaS$_2$, as well as samples S1 and S2. This data is fit to the linearized gap equation, with $\Delta^\parallel_\textrm{so}$ and $\Delta^\perp_\textrm{so}$ as fitting parameters. In agreement with the expected outcome in prior reports\cite{NbSe2Ising,TaS2}, $H^\parallel_\textrm{c2}$ for NbSe$_2$ and TaS$_2$ follows square root temperature dependence over the entire measurable range, and yields $\Delta^\parallel_\textrm{so} \sim 0$. For S1, the 
measured $H^\parallel_\textrm{c2}$ clearly falls below the square-root dependence at low T, and the tilted Ising model provides an excellent fit to the data over the entire field and temperature range. We find $\Delta^\perp_\textrm{so} \sim 2.34$ meV, with $\Delta^\parallel_\textrm{so}$/$\Delta^\perp_\textrm{so}$, of $\sim$15\%, in good agreement with the splitting from the DFT calculated electronic bandstructure shown in Fig. 1b. As a check, we see that the Klemm-Luther-Beasley (KLB) model, which applies to superconductivity in the dirty limit with strong SOS, overestimates $H^\parallel_\textrm{c2}$ at low $T$. This is in agreement with estimations of spin-orbit scattering times indicating that S1 and S2 do not meet the criteria for the dirty limit (see Supplementary for details). For S2 (Fig. 4c), we find $\Delta^\perp_\textrm{so} \sim 1.5$ meV, with $\Delta^\parallel_\textrm{so}$/$\Delta^\perp_\textrm{so}$ of $\sim$4.6\%, a third as much as in S1. \\
\indent The excellent fit to the data, and observed variation between the samples with presumably different $\phi$, provides strong confirmation of the model. However, open issues remain. Most importantly, the change in the ratio of $H^\parallel_\textrm{c2}/H_\textrm{p}$ between S1 and S2 is less than expected, as can be seen from Fig. 4a, given the difference of $\Delta^\parallel_\textrm{so}$/$\Delta^\perp_\textrm{so}$ between the two samples. This suggests that other additional mechanisms may be at play in limiting $H^\parallel_\textrm{c2}$. One possibility is the recently proposed effect of spin-orbit-parity coupling (SOPC) enhanced upper critical fields\cite{xie2020spin}. Determination of the dominant mechanism for enhanced $H^\parallel_\textrm{c2}$ between SOPC and SOC remains to be explored. Other effects, such as sample quality and doping level, may also contribute to the different SOC parameters found in S1 and S2. Full measurement of the angle-dependent response at high magnetic fields will greatly clarify these issues. \\
%
%
\indent The telluride family of TMDs remains poorly explored, and rich phenomena should emerge now that reliable and clean fabrication processes are available for the study of air-sensitive monolayer films. The demonstration here of strongly enhanced, gate-tunable superconductivity in 1L-MoTe$_2$ should motivate significant future studies to understand both the mechanism for the observed enhancement and to confirm the proposed \textit{tilted} Ising SOC, which remains an experimental challenge due to the need for in-plane rotation in magnetic fields in excess of 25 T and temperatures below 1 K. Additionally, the electronic structure of MoTe$_2$ is highly sensitive to external inputs like strain or electric fields\cite{ReedEDope,ReedStrain,shi_symmetry_2019, OriginMoTe2}, allowing study of how the superconducting phase may be modified by varying the SOC or inducing topological insulator states. Finally, we note that recent investigations into bulk $T_d$-MoTe$_2$ have revealed the existence of superconducting edge currents\cite{wang2020evidence}. While this result remains to be shown in few-layer MoTe$_2$, electrostatic control over carriers and higher $T_\textrm{c}$ over that of the bulk for such a state could provide a platform for quantum logic devices based on topological protection without the necessity of a dilution refrigerator. 
%
\begin{acknowledgments}
We would like to acknowledge A. Benyamini and E. Telford for fruitful discussions. This work was primarily supported by the NSF MRSEC program through Columbia in the Center for Precision Assembly of Superstratic and Superatomic Solids (DMR-1420634).  H.W. and X.Q. acknowledge the funding support from NSF DMR-1753054. L.B. acknowledges support from NSF DMR-1807969. N.F.Q.Y. and L.F. are supported by DOE Office of Basic Energy Sciences, Division of Materials Sciences and Engineering (DE-SC0010526). L.F. is partly supported by the David and Lucile Packard Foundation. A portion of this work was performed at the National High Magnetic Field Laboratory, which is supported by the NSF Cooperative Agreement No. DMR-1644779 and the State of Florida. Portions of this research were conducted with the advanced computing resources provided by Texas A\&M High Performance Research Computing.\\
\indent D.R. and A.J. contributed equally to this study.
\end{acknowledgments}

\bibliography{PRL}
\end{document}


\maketitle

\begin{affiliations}
 \item Department of Mechanical Engineering, Columbia University, New York, NY, USA
 \item Department of Physics, Columbia University, New York, NY, USA
 \item Department of Physics, Massachusetts Institute of Technology, Cambridge, MA, USA
 \item Department of Materials Science and Engineering, Texas A\&M University, College Station, Texas, USA
 \item Department of Physics, Florida State University, Tallahassee, FL, USA
 \item National High Magnetic Field Laboratory, Florida State University, Tallahassee, FL, USA
 \item National Institute for Materials Science, 1-1 Namiki, Tsukuba 305-0044, Japan
 \item Department of Applied Physics and Applied Mathematics, Columbia University, New York, NY, USA
\item[$*$] These authors contributed equally to this work.
 
\end{affiliations}

\maketitle
\newpage
\section*{Single Crystal Growth}
\indent High quality single crystals were prepared by combining molybdenum powder (99.999\% ) and tellurium lumps (99.9999+\%) in a ratio of 1:20 in a quartz ampoule. The ampoules were subsequently sealed under vacuum ($\sim 5 \times 10^{-6}$ Torr). The reagents were then heated to 1100 $^{\circ}$C within 24 h and dwelled at this temperature for 24 h before being cooled to 880 $^{\circ}$C over 400 h. At 880 $^{\circ}$C the tellurium flux was decanted in a centrifuge and the samples was quenched in air. The subsequently obtained 1T'-MoTe$_2$ single crystals were annealed at 425 $^{\circ}$C with a 200 $^{\circ}$C gradient for 48 h to remove any residual tellurium. As a quality check, previous batches following this recipe have yielded bulk crystals with a residual resistivity ratio of 700 to 2000. 
\section*{Fabrication}
\textbf{Via Devices:} For sample fabrication 20-30 nm thick \textit{h}-BN flake is first exfoliated onto SiO$_2$, and holes are etched using a SF$_6$ or CHF$_3$/O$_2$ plasma mix. A second, larger diameter hole is then opened through a PMMA mask and then Pd/Au, (20/50 nm), or Au (50 nm) is deposited into the holes, forming an embedded contact in the \textit{h}-BN flake that has an overlapping portion, henceforth referred to as via contact\cite{ViaEvan}. This is then picked up by a dry transfer method using polypropylene carbonate (PPC) as the pickup polymer\cite{1Dcontact}. \\
\indent Monolayer $T_d$-MoTe$_2$ flakes are exfoliated onto PDMS inside a nitrogen-filled glovebox., identified optically, and subsequently picked up from PDMS using the already picked up via contacts. Stacks are then placed onto a previously Ar-O$_2$ annealed\cite{CleanBN}, 20-30 nm thick, \textit{h}-BN crystal. After encapsulation, the devices are removed from the glovebox, the PPC removed by rinsing in chloroform or acetone, and etched using O$_2$-CHF$_3$ plasma. Contacts were fabricated through conventional e-beam lithography techniques (EBL) and Ti/Au or Cr/Au, 5/75 nm, was deposited to make contact to the via contacts, with the overlapping region being larger to insure that any gap caused during the transfer process is filled, preventing further exposure of the sample to air. Figure 1 depicts the total process flow.\\ 
\noindent\textbf{Prepatterned Contact Devices:} While prepatterned contacts had less yield over via devices for monolayer, the yield for multilayer devices using prepatterned contacts is dramatically improved over that of via contacts. Using 12 nm Au, or AuPd, prepatterned contacts on 20-30 nm \textit{h}-BN we were able to achieve nearly 100\% yield of successful devices when proper care was taken into consideration post-stacking. First, metal backgates are patterned on to 285 nm SiO$_2$ using standard EBL techniques and a bilayer - PMMA 495A4/PMMA 950A2 resist mask. After patterning, 10 nm of AuPd was deposited via e-beam deposition, followed by 300 $^\circ$C vacuum annealing, and low power O$_2$ plasma to remove any left over residue on the prepatterned metal backgates. 20-30 nm \textit{h}-BN is then transferred on to the metal backgates using the aforementioned dry stacking method. Subsequently, contacts are patterned on to the \textit{h}-BN using standard EBL and 12 nm or either Au or AuPd are deposited via e-beam deposition. After liftoff in acetone, contacts are cleaned by contact AFM in a Dimension Icon Bruker, with deflection ranging from 0.3 - 0.6 V. All subsequent steps are the same as described in "Via Devices." Overdoped samples (i.e. increasing $T_\text{c}$ with positive $V_\text{bg}$) were typically produced when only Si/SiO$_2$/\textit{h}-BN was used as a backgate, i.e. sample S2 from the main text. However, using a local graphite backgate produced the overall highest $T_\text{c}$ and shows a pronounced peak in resistivity at low backgate bias, see sample S1 in the main text.\\
\section*{Transport Measurements}
\indent Supplementary Figure 2a depicts the temperature dependence for four different thicknesses of encapsulated MoTe$_2$. In the bulk a clear transition can be see at 240 K, however this transition is absent in the few-layer samples. In agreement with previous reports\cite{TsenMoTe2}, this is due to a gradual transition from the $T'$ phase to the $T_d$ phase as one decreases interlayer coupling (i.e. reduces the number of layers). We also begin to observe the onset of superconductivity for in a monolayer sample. In Supplementary Fig. 2b. we plot the resistivity versus temperature for samples S1 and S2. Sample S2 displays a resistivity (multiplied by a factor of 10) of 76 $\Omega$. For each sample one can calculate the mean free path using either the Drude model: $\sigma = ne^2l_\text{Drude}/m^*v_\text{F}$, where $\sigma$ is the conductivity, $n$ is the carrier density, $e$ the charge of an electron, $m^*$ the effective mass (taken from the bandstructure calculations as 0.87 $m_\text{e}$, the electron bare mass) or through a similar formalism which takes into account the degeneracy of valley and spin: $l_\text{deg} = h\sigma/(e^2\sqrt{g_sg_v\pi n})$, here $g_s=g_v=2$ are the spin and valley degeneracy\cite{CobdenWTe2}, respectively. However, as monolayer $T_d$-MoTe$_2$ is semimetallic, the exact value of the carrier density is difficult to extract experimentally \textemdash showing unrealistically large changes in carrier density as a function of gate induced carrier density (see Supplementary Fig. 3). This is amplified by the difficulty of producing samples with good Hall geometry, as these samples can not be etched into the channel, otherwise the sample degrades drastically.\\ 
\indent Given these complications, we evaluate the mean free path assuming a carrier density, $n_\text{DFT}$, extracted from the density functional theory (DFT) calculated Fermi surfaces (FS), which totals to $\sim2.4\times 10^{13}$/cm$^2$, where a single hole pocket has a higher carrier concentration of $1.2\times10^{13}$ cm$^{-2}$ as compared to that of the electrons ($0.6\times10^{13} $cm$^{-2}$ each) . While we have not verified directly the carrier density of the monolayer, the validity of this assumption can be verified through transport measurements of bilayer, and trilayer samples where Shubnikov-de Haas (SdH) oscillations are visible in the magnetoresistance (see Supplementary Fig. 4). Shown in Supplementary Fig. 4b, both the bilayer and trilayer samples exhibit superlinear behavior for the normal state resistance as a function of the applied magnetic field. Dashed lines are fits to the function: $R_\textrm{xx} = R_\textrm{0}+\gamma B^\alpha$, where $\gamma$ represents a combination of sample geometry, mobility, and carrier density and $R_\textrm{0}$ is the typical $(l/w)[ne(\mu_\textrm{e}+\mu_\textrm{h})]$. For perfect compensation, $\alpha$ is expected to be 2, whereas for single dominant carrier $\alpha = 0$.Shown in Supplementary Fig. 4b, fitting to the trilayer sample yields a value of 1.71, and to the bilayer 1.56. Deviations from 2 can either indicate a slightly larger carrier population for either holes or electrons,  an anisotropic FS, nontrivial FS topology, or disorder\cite{mitra2019quadratic}. In this case, it's likely a combination of disorder and slight offset from perfect compensation given the trilayer has a value closer to 2 than that of the bilayer and concomitantly has a higher mobility, which is estimated from the onset of the SdH oscillations (950 cm$^2$/Vs for the trilayer, 850 cm$^2$/Vs for the bilayer). Background subtraction, using a third degree polynomial, of the magnetoresistance yields an SdH signal in both the bilayer and trilayer with a frequency of 412-449 T. Using the frequency of the oscillations with respect to the inverse of the magnetic field ($B_\textrm{F}$) in combination with the Onsager relation for a magnetic field, $B_\textrm{F} = \phi_\textrm{0}\textrm{A}/2\pi^2$, where $\phi_\textrm{0}$ is the magnetic flux quantum, and assuming a 2D, circular Fermi surface, $\textrm{A} = \pi k_\textrm{F}^2$, $k_\textrm{F} = \sqrt{2\pi n}$ yields a carrier density of $\sim$2$\times 10^{13}$ cm$^{-2}$ - in reasonably good agreement with the DFT calculated carrier density. This also implies that the frequency we are measuring is likely due to the larger hole pocket.\\
%
\indent Using the aforementioned carrier density, the mean free paths are calculated for samples S1 and S2 and summarized in Table 1. For S2, the clean limit is easily achieved as the mean free path (307 nm) is twenty times larger than that of the coherence length (14-16 nm). S1, on the other hand, has a much smaller mean free path on the order of the coherence length at zero gate voltage. However, as indicated by the unrealistic spin-orbit scattering times (see table 2), we are still able to fit our spin-orbit coupling model for $H_\textrm{c2}^\parallel$ vs $T$ (see below) and extract values for the spin-orbit coupling strengths in- and out-of-plane in general agreement with sample S2.\\
%
\section*{Fitting of experimental upper critical field on 1L-MoTe$_2$}
\textbf{Klemm-Luther-Beasley:} The KLB model is defined by a fit to  $\ln({T/T_c})+\Psi(\frac{1}{2}+\frac{3\tau_\textrm{SO}\mu^2_B H^2_\textrm{c2}}{4\pi\hbar k_BT})-\Psi(\frac{1}{2}) = 0$, where $\Psi$ is the digamma function, $\tau_{SO}$ the spin-orbit scattering time,  and $\mu_B$ is the Bohr magneton.
\textbf{With in- and out-of-plane SOC:} For a paramagnetic limiting field at finite temperature, $T$, we consider the linearized gap equation:\\
$\log(T/T_\textrm{c})+\left \langle w_{\bm{k},+}\Phi(\rho_{\bm{k},+})+w_{\bm{k},-}\Phi(\rho_{\bm{k},-}) = 0 \right \rangle_\text{FS}$, with\\ 
$w_\pm = \frac{1}{2}\left [ 1 \pm \frac{\mid{\bm{g}}\mid^2-\mid\mu \bm{B}\mid^2}{\mid\bm{g}+\mu\bm{B}\mid\cdot\mid\bm{g}-\mu\bm{B}\mid} \right ]$, $\rho_\pm = \frac{\mid\bm{g}+\mu\bm{B}\mid\mp\mid\bm{g}-\mu\bm{B}\mid}{2\pi T}$, $\Phi(x) = Re\left \{ \psi\left (\frac{1+ix}{2} \right ) -\psi\left (\frac{1}{2} \right ) \right \}$.\\ 
Where $\bm{g}$ is the spin-orbit field, $\bm{B}$ is the magnetic field, $\mu$ is the electron magnetic moment, and $\left \langle ... \right \rangle_\text{FS}$ is the Fermi surface average: 
$\left\langle f\bm{_k} \right\rangle_\textrm{FS}$ 
$ \equiv \frac{1}{N_0}\int_{\xi\bm{_k}=0}$ $\frac{d^2\bm{k}}{(2\pi)^2}f\bm{_k}$, where $\xi\bm{_k}$ is the kinetic energy and $\xi\bm{_k} = 0 $ is the Fermi surface. We can then fit the upper critical field data using $\bm{g_k} = (\alpha_yk_y,\alpha_xk_x,\alpha_Ik_y)$. For an in-plane field $\bm{B} = B(\cos{\phi}\bm{\hat{x}}+\sin{\phi}\bm{\hat{y}})$ we have:\\
$\left\langle \left | \bm{g_k}\times\bm{\hat{B}} \right |^2 \right\rangle^{1/2}_\text{FS} = \sqrt{(\alpha_y^2+\alpha_I^2\sin{^2\phi})K_y^2+\alpha_x^2K_x^2\cos{^2\phi}}$,\\
 and $\left\langle \left | \bm{g_k}\cdot\bm{\hat{B}} \right |^2 \right\rangle^{1/2}_\text{FS} = \sqrt{\alpha_y^2K_y^2\cos{^2\phi}+\alpha_x^2K_x^2\sin{^2\phi}}$, $K_{x,y} \equiv \left\langle k^2_{x,y} \right\rangle^{1/2}_\text{FS}$. In the main text, we have defined $\Delta_\text{so}^\parallel/\Delta_\text{so}^\perp = \frac{\left\langle \left | \bm{g_k}\cdot\bm{\hat{B}} \right |^2 \right\rangle^{1/2}_\text{FS}}{\left\langle \left | \bm{g_k}\times\bm{\hat{B}} \right |^2 \right\rangle^{1/2}_\text{FS}} $\\ 
Using this formalism, we fit samples one and two from the main text to reconfirm the differing spin-orbit coupling, shown in Supplementary Fig. 6. From Supplementary Fig 6a., S1 shows a considerable in plane component of spin-orbit coupling (SOC). In order to evaluate the accuracy of these fits, we attempt to fit several other values for the spin-orbit coupling strengths in- and out-of-plane. As shown, small deviations in the overall in-plane spin-orbit coupling strength show considerable differences, from overestimation at low in-plane values, to an uncharacteristic rollover and underestimation at substantially higher in-plane SOC. As such, our best fit gives a relative coupling strength of in-plane SOC as compared to out-of-plane SOC, $\Delta_\text{SOC}^\parallel/\Delta_\text{SOC}^\perp$, of 15.3\%. However, in the case of samples where $H_\text{c2}^\parallel$ is only measured down to values of $T/T_\text{c} \sim .5$ with larger out-of-plane SOC the uncertainty in the fitting becomes substantially larger. In sample S2 (see Supplementary Fig. 6b.), where this is indeed the case, we find a relative coupling strength for in-plane SOC as compared to out-of-plane SOC, $\Delta_\text{SOC}^\parallel/\Delta_\text{SOC}^\perp$ between 0 to 7.5\% fits reasonably well, but values larger than 7.5\% repeatedly underestimate the experimental values between $T = 2.0$ K and $T = 3.5$ K. Overall, the two samples are in qualitative agreement with the expectation of an anisotropic in-plane magnetic field dependence as described in the main text.  
\section*{First-principles electronic structure calculations}
The ground-state crystal structure of monolayer MoTe$_2$ was obtained from first-principles density functional theory (DFT) calculations as implemented in the Vienna Ab initio Simulation Package (VASP) using the plane wave basis\cite{kresse_efficient_1996} with an energy cutoff of 400 eV and the projector-augmented wave method\cite{blochl_projector_1994}. We adopted the Perdew-Burke-Ernzerhof’s form of exchange-correlation functional within the generalized-gradient approximation and the Monkhorst-Pack k-point sampling of 8$\times$16$\times$1 for the integration over the first Brillouin zone as well as a vacuum slab of 16 $\text{\AA}$ along the z direction to reduce image interaction under the periodic boundary condition in the calculation. The monolayer crystal structure was fully relaxed with a maximum residual force of less than 0.01 eV/\AA. To investigate the detailed electronic structure, we performed DFT calculations with a hybrid HSE06 exchange-correlation functional\cite{heyd_hybrid_2003} and transformed the DFT Kohn-Sham eigenstates into a set of highly localized quasiatomic orbitals and corresponding first-principles tight-binding Hamiltonian\cite{qian_quasiatomic_2008}. Using the effective Hamiltonian we computed the corresponding fine electronic band structure and spin texture in the 2-D Brillouin zone. An electric field of 0.01 V/$\text{\AA}$ was applied along the z direction which breaks the C$_2$ two-fold rotation symmetry, thus breaking the inversion symmetry and leads to spin splitting which qualitatively matches the experimental data.
\newpage
%
\begin{figure*}[t] 
    \centering
	\includegraphics[width=\linewidth]{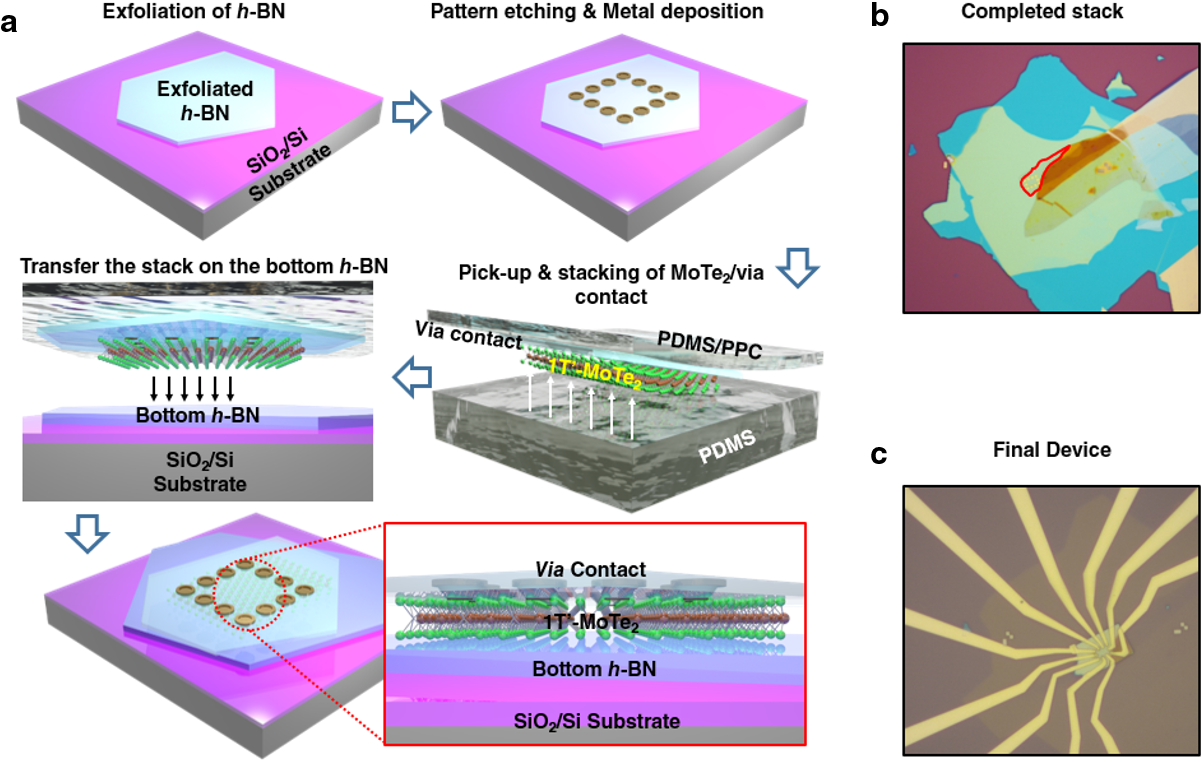}
	\caption*{\textbf{Supplementary Figure 1 $|$ Device Fabrication. a}, Depiction of the initial via fabrication and stacking sequence. \textbf{b}, Optical image of a completed stack, monolayer portion is outlined in red. \textbf{c}, Optical image of a completed device.}
\label{Fig. S1}
\end{figure*}
\newpage
%
\begin{figure*}[t] 
    \centering
	\includegraphics[width=\linewidth]{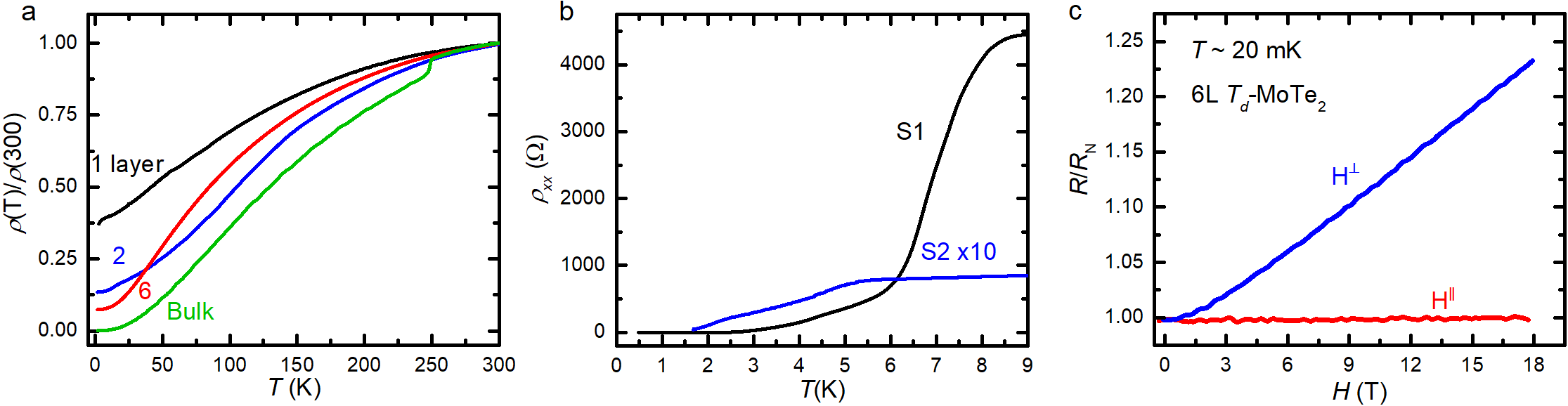}
	\caption*{\textbf{Supplementary Figure 2 $|$ Temperature dependence of normalized resistivity for various thicknesses of $T_d$-MoTe$_2$.}\textbf{a} Temperature dependence of the normalized resistivity for bulk, 6-layer, bilayer, and monolayer samples. Notice the lack of a phase transition in the few-layered samples. \textbf{b} Low temperature $T$-dependence of two monolayer samples, S1 and S2, both showing enhanced $T_\textrm{c}$. S1 is a graphite backgated sample, whereas S2 utilizes a silicon backgate. \textbf{c} Parallel and perpendicular magnetic field dependence for a 6-layer sample of $T_d$-MoTe$_2$ at $T = 20$ mK.}
	
\label{Fig. S2}
\end{figure*}
\clearpage
%
\begin{figure*}[t] 
    \centering
	\includegraphics[width=\linewidth]{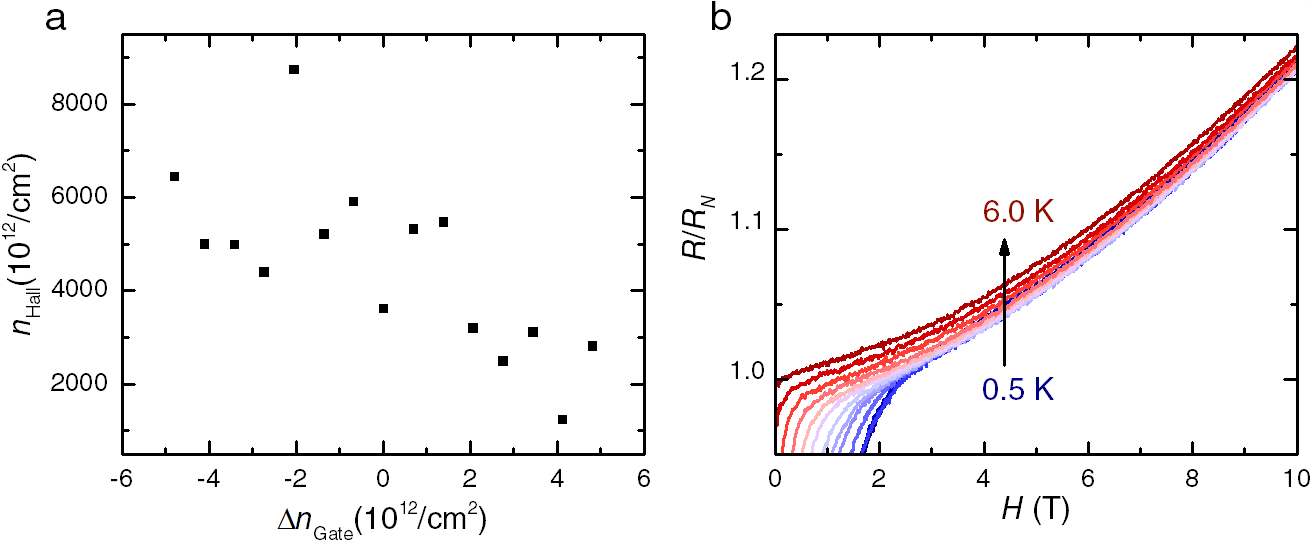}
	\caption*{\textbf{Supplementary Figure 3 $|$ Semimetallic behavior. a,} Experimental single carrier density, $n_\text{Hall}$, vs. gate induced carrier density, $\Delta n_\text{Gate}$. \textbf{b}, Normalized Magnetoresistance as a function of field above $H^\perp_\text{c2}$ for sample S2 and for $V_\textrm{bg} = 0$ V. Notice the nonlinear behavior.}
	
\label{Fig. S3}
\end{figure*}
\newpage
\clearpage
%
\begin{figure*}[t] 
    \centering
	\includegraphics[width=\linewidth]{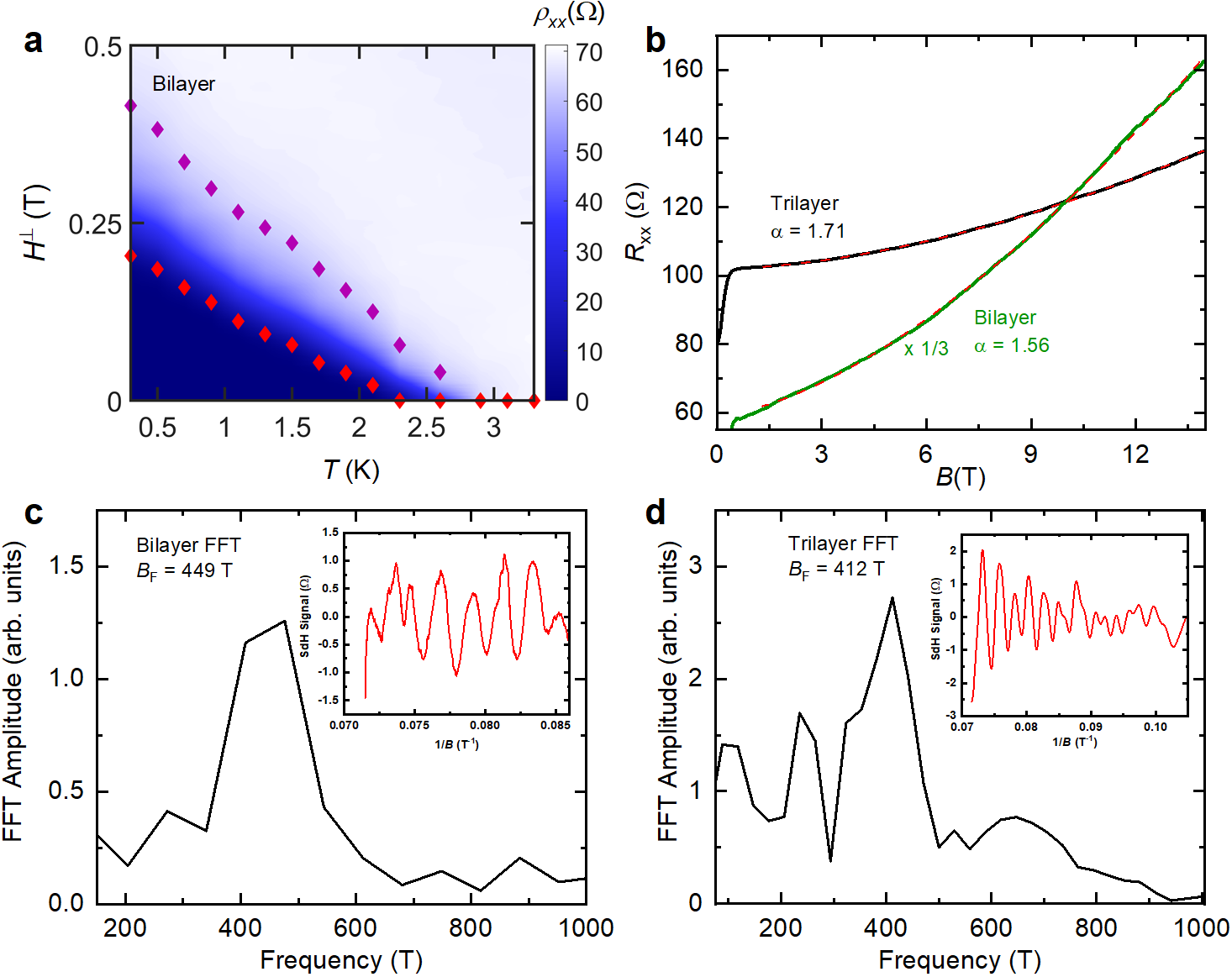}
	\caption*{\textbf{Supplementary Figure 4 $|$ Bilayer and Trilayer Behavior. a,} Perpendicular magnetic field vs temperature phase diagram for a bilayer sample with $T_\textrm{c} = 2.5$ K, where a coherence length of $\sim$30 nm can be extracted. \textbf{b} $R_\textrm{xx}$ as a function of the magnetic field in the perpendicular direction, dashed lines are fits the polynomial function described in the text. \textbf{c,d} The fast Fourier transform of the SdH signal as extracted from (b) for the bilayer and trilayer sample, respectively. Inset c,d: the SdH signals for both the bilayer and trilayer samples, respectively}
	
\label{Fig. S4}
\end{figure*}
\newpage
%
\clearpage
\begin{figure*}[t] 
    \centering
	\includegraphics[width=.5\linewidth]{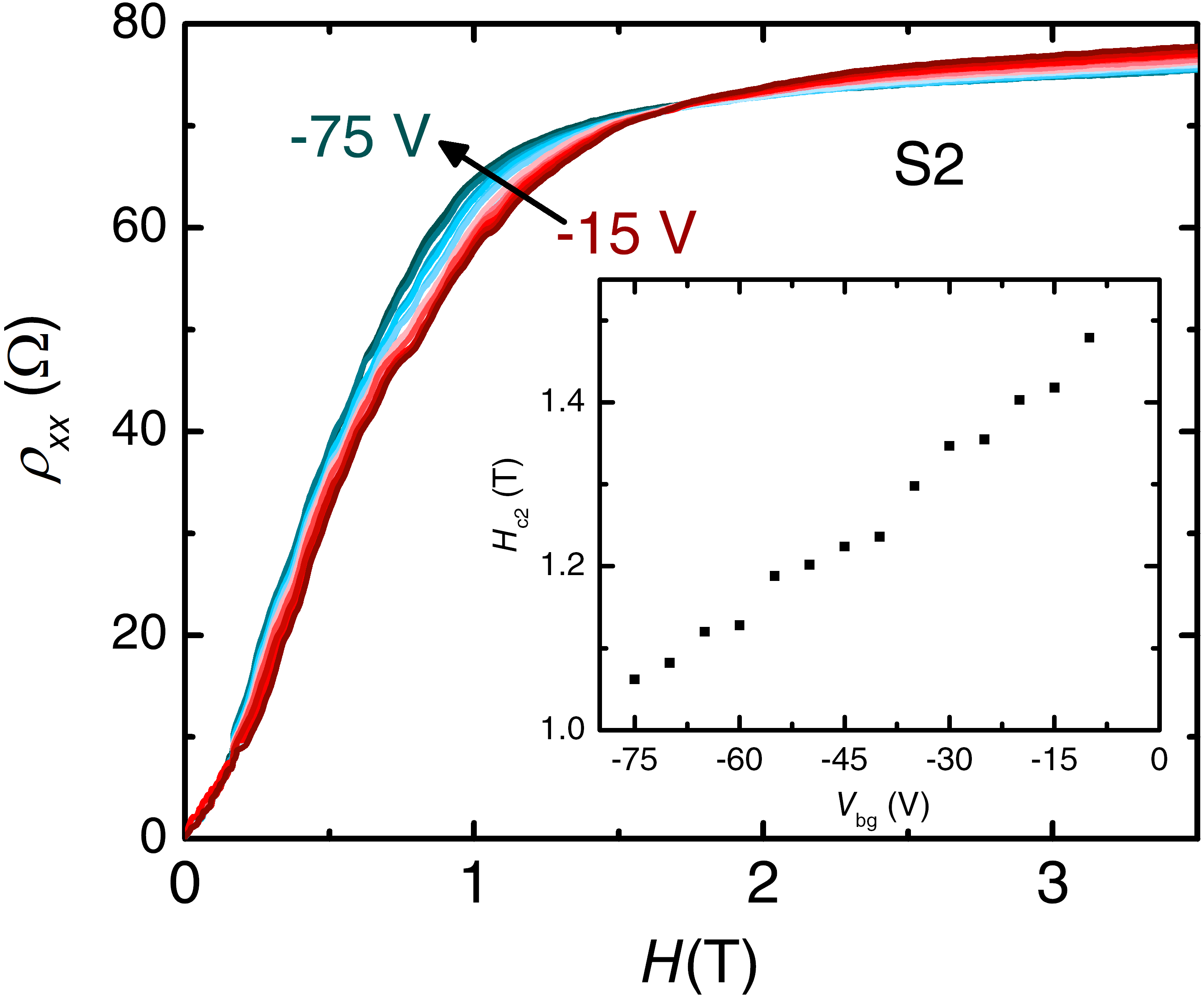}
	\caption*{\textbf{Supplementary Figure 5 $|$ $\Delta n_\text{Gate}$.} Resistivity vs field and for several values of gate voltage applied to sample S2. Notice that the sheet resistance in the normal state increases monotonically as a function of increasing gate voltage. Inset: $H^\perp_\text{c2}$ monotonically increasing as a function of the backgate voltage.}
	
\label{Fig. S5}

\end{figure*}
\clearpage
%
\begin{figure*}[t] 
    \centering
	\includegraphics[width=.75\linewidth]{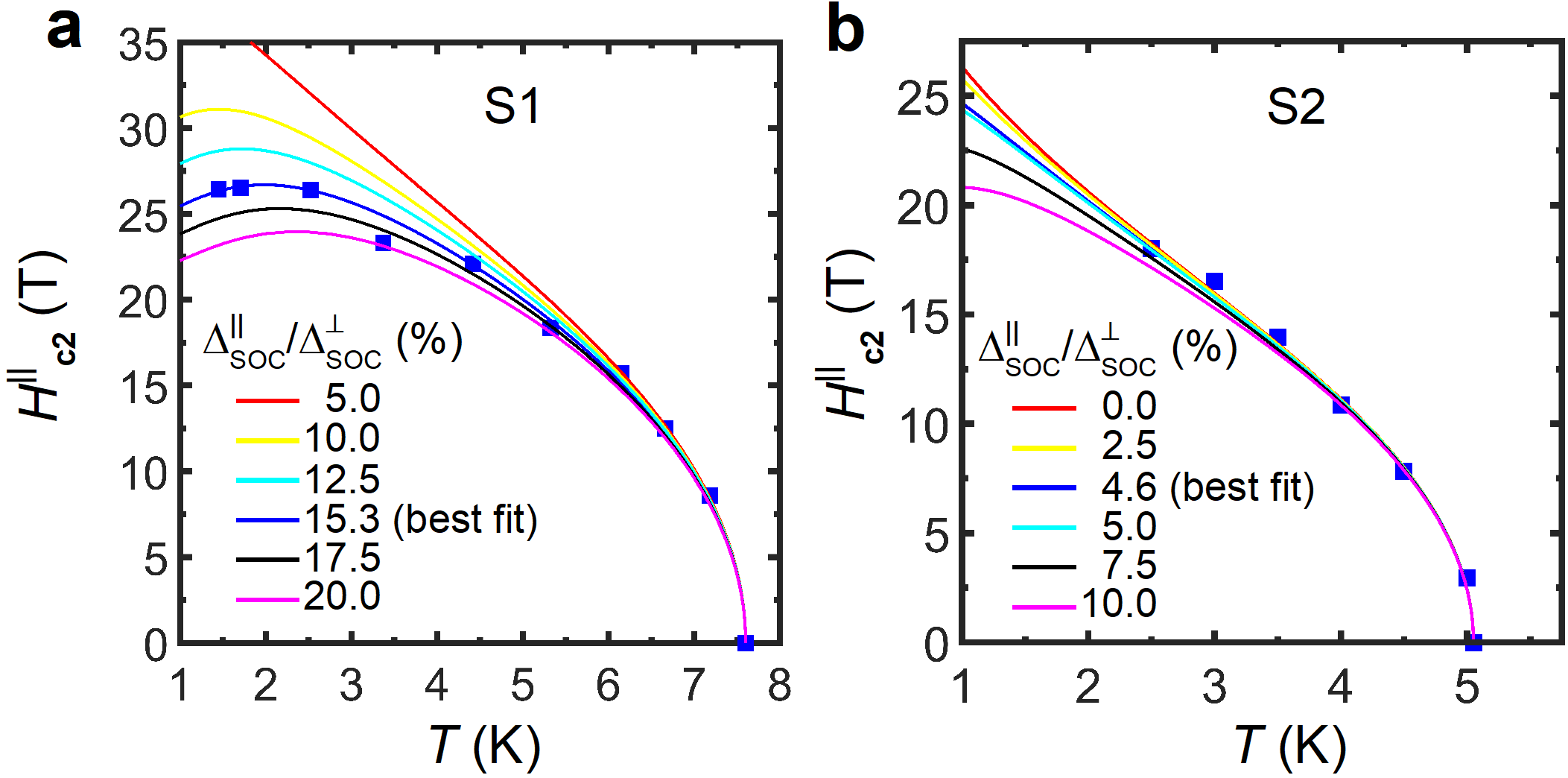}
	\caption*{\textbf{Supplementary Figure 6 $|$ Spin-orbit coupling fittings. a-b}  Measured $H^\parallel_\textrm{c2}$ for fields parallel to the \emph{ab}-plane for Samples 1, and 2 with fits to several other relative spin-orbit coupling strengths for reference. Blue squares are experimental data, solid lines are theoretical fits using the formalism described in the text.}
	
\label{Fig. S6}
\end{figure*}
\clearpage
%
\begin{figure*}[t] 
    \centering
	\includegraphics[width=.5\linewidth]{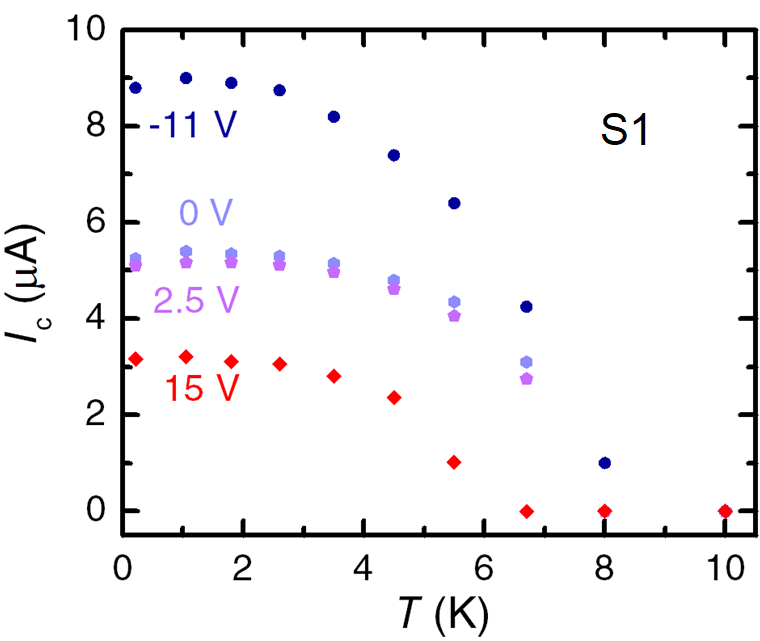}
	\caption*{\textbf{Supplementary Figure 7 $|$ Critical current. a} Critical current, $I_\textrm{c}$ as a function of temperature and varying gate voltage for sample S1 described in the main text. As in the case of the critical temperature, $I_\textrm{c}$ increases monotonically as the gate voltage is decreases.}
	
\label{Fig. S7}
\end{figure*}
\clearpage
%
\begin{table}
\centering
\caption*{ \textbf{Table 1. Summary of the mean free paths for each sample.} The mean free paths are either derived through the Drude model for the value of $n$ given by the SdH oscillations ($n \sim 2\times10^{13}$/cm$^2$) or by the model which takes into account valley and spin degeneracy given in the text.} 

\begin{tabular}{|p{.45 cm}||p{1.1 cm}|p{1.1 cm}|p{1.25 cm}|p{1.9 cm}|p{2.4 cm}|p{1.65 cm}|p{2.4 cm}|}
   \hline & $T_\text{c}$ (K) & $\xi_0 (nm)$ & $\rho_{xx}$ ($\Omega$) & $l_\text{Drude}$ (nm) 
   & $l_\text{deg}$ (nm) \\ \hline 
S1 & 7.6 & 7.5 & 3560 & 6.73 & 4.76 \\ \hline 
S2 & 5 & 13.9 & 78 & 307 & 217 \\ \hline 
\end{tabular}

\end{table}
\clearpage
\begin{table}
\centering
\caption*{ \textbf{Table 2. Summary of scattering rates.} The scattering rates are either derived through the Drude model for the calculated value of $v_\text{F}$ and $n$ from the theoretical electronic band structure ($n \sim 2\times10^{13}$/cm$^2$) or by assuming an upper limit carrier density of $n = 1\times10^{14}$/cm$^2$ and deriving $v_\text{F}$ assuming a 2-D circular Fermi surface, i.e. $v_\text{F} = \hbar \sqrt{2\pi n}/m^*$ and using the values of the mean free path outlined in Table 1. The values for the spin-orbit coupling are taken from the best fits outlined in Fig. S5.} 

\begin{tabular}{|p{.45 cm}||p{1.67 cm}|p{1.67 cm}|p{1.5 cm}|p{1 cm}|p{1 cm}|}
   \hline & $\tau_\text{Drude}$ (fs)
   
   & $\tau_\text{deg}$ (fs) 
   
   & $\tau_\text{SOS}$ (fs)
   
   & $\Delta^\parallel$ 
   
   (meV) & $\Delta^\perp$
   
   (meV)
\\ \hline
S1 & 54 & 10 & 88-101 & 0.358 & 2.34\\ \hline 
S2 & 2463 & 1742 & 89-150 & 0.057 & 1.51\\ \hline 
\end{tabular}
\end{table}

\clearpage
%
%
\clearpage
\section*{References}
\bibliographystyle{naturemag}
\bibliography{PRL}